\documentclass[twocolumn,aps,amssymb,superscriptaddress, showpacs]{revtex4}
\usepackage{graphicx}

\begin{document}

\title{Spin-phonon coupling in ZnCr$_{2}$Se$_{4}$}

\author{T.~Rudolf}
\author{Ch.~Kant}
\author{F.~Mayr}
\author{J.~Hemberger}
\affiliation{Experimental Physics V, Center for Electronic
Correlations and Magnetism, University of Augsburg,
D-86135~Augsburg, Germany}

\author{V.~Tsurkan}
\affiliation{Experimental Physics V, Center for Electronic
Correlations and Magnetism, University of Augsburg,
D-86135~Augsburg, Germany} \affiliation{Institute of Applied
Physics, Academy of Sciences of Moldova, MD-2028~Chisinau, Republic
of Moldova}

\author{A.~Loidl}
\affiliation{Experimental Physics V, Center for Electronic
Correlations and Magnetism, University of Augsburg,
D-86135~Augsburg, Germany}

\date{\today}

\begin{abstract}
Spin-phonon coupling and magnetodielectric effects of ZnCr$_2$Se$_4$
were investigated by means of infrared (IR) spectroscopy as a
function of temperature and magnetic field. ZnCr$_2$Se$_4$ is
dominated by ferromagnetic exchange but undergoes antiferromagnetic
order at $T_\mathrm{N} = 21$~K. In the magnetically ordered phase
the low-frequency IR active phonon splits, indicating strong dynamic
anisotropy via magnetic exchange interactions. Antiferromagnetic
order and concomitantly the phonon splitting is wiped out by
external magnetic fields. Hence, ZnCr$_2$Se$_4$ is a prominent
example of a spin-driven Jahn-Teller effect which can be fully
suppressed in external magnetic fields.
\end{abstract}



\maketitle

Recently spin-phonon coupling (SPC) in materials with strong
electronic correlations gained considerable attention. The onset of
low-temperature magnetic order of spins residing on a pyrochlore
lattice in geometrically frustrated spinel oxides has been explained
with the concept of a spin-driven Jahn-Teller effect
\cite{Yamashita,Tchernyshyov}. In ZnCr$_2$S$_4$ complex magnetic
order results from strong bond frustration characterized by
ferromagnetic (FM) and antiferromagnetic (AFM) exchange interactions
of almost equal strength \cite{Hemberger97}. All these compounds
have in common that the onset of AFM order is only accompanied by
small structural distortions, hardly detectable by standard
diffraction techniques, but induces a strong splitting of specific
phonon modes. This phonon splitting has been experimentally verified
in detail in ZnCr$_2$O$_4$ \cite{Sushkov} and ZnCr$_2$S$_4$
\cite{Hemberger97} by optical spectroscopy. It has been explained by
ab-initio approaches providing evidence that the dynamic anisotropy
of the phonon modes is induced by the magnetic exchange alone and
can be fully decoupled from lattice distortions
\cite{Massidda,Fennie}. The idea that spin-phonon interactions and
the onset of magnetic order strongly influence the phonon modes has
been outlined earlier by Baltensperger and Helman
\cite{Baltensperger} and by Br\"{u}esch and D'Ambrogio \cite{Brueesch}.
SPC also is of prime importance in a number of multiferroic
compounds and has been demonstrated for BiFeO$_3$ \cite{Haumont} and
TbMn$_2$O$_5$ \cite{Aguilar}. In multiferroics the strong coupling
of phonons and magnons even creates a new class of excitations,
namely electromagnons, which are spin excitations observable by ac
electric fields \cite{Pimenov}.

In the normal spinel ZnCr$_2$Se$_4$ \cite{Hahn} the Cr spins reside
on the B-site pyrochlore lattice. The paramagnetic moment is close
to the spin-only value of Cr$^{3+}$ and the susceptibility reveals a
large positive Curie-Weiss (CW) temperature indicative for dominant
FM exchange \cite{Lotgering}. Even so, ZnCr$_2$Se$_4$ reveals AFM
order below 21~K. For $T < T_\mathrm{N}$ the spin structure is
characterised by ferromagnetic (001) planes with a turn angle of the
spin direction of 42$^\circ$ between neighbouring planes
\cite{Plumier,Akimitsu}. The magnetic moment of the Cr$^{3+}$ ions
in the spiral state is $1.71~\mu_B$, remarkably smaller than the
paramagnetic moment. This large reduction of the ordered moment has
not been explained so far. The magnetic phase transition is
accompanied by a small tetragonal distortion with $c/a = 0.9992$ at
4.2~K \cite{Kleinberger}. The structural modulation in the vicinity
of the antiferromagnetic phase transition has been investigated in
detail by Hidaka and coworkers \cite{Hidaka} by neutron and
synchrotron radiation and it has been shown that magnetic and
structural orders are strongly influenced already by moderate
magnetic fields. It is interesting to note that also magnetoelectric
effects have been reported for ZnCr$_2$Se$_4$ and that the field
dependence of the observed effects has been qualitatively explained
by Dzyaloshinskii-Moriya type interactions \cite{Siratori}. Finally,
very recently strong spin-phonon coupling has been evidenced by the
observation of giant negative thermal expansion close to
$T_\mathrm{N}$ and it has been shown that antiferromagnetic order
can be shifted to 0~K in external magnetic fields of 6~T
\cite{Hemberger06}.

Here we document SPC in ZnCr$_2$Se$_4$, which is dominated by strong
FM exchange, but orders antiferromagnetically at $T_\mathrm{N} =
21$~K. ZnCr$_2$Se$_4$ belongs to a group of chromium spinels with
competing FM and AFM interactions, in which the relative strength of
both varies considerably with the lattice constants \cite{Baltzer}.
In ZnCr$_2$O$_4$ strong direct AFM exchange of Cr-spins on a
pyrochlore lattice determines the complex spin order at low
temperatures \cite{Lee}. In CdCr$_2$S$_4$ ferromagnetic 90$^\circ$
Cr-S-Cr exchange is responsible for a purely ferromagnetic ground
state \cite{Hemberger05}. In ZnCr$_2$S$_4$ next-nearest neighbour
AFM exchange and 90$^\circ$ FM exchange are of equal strength,
leading to bond frustration \cite{Hemberger97}. In ZnCr$_2$Se$_4$
the strong FM exchange governs the ferromagnetic order within the
(001) planes but an obviously weak AFM interplane coupling is
responsible for the AFM ground state and a concomitant splitting of
a phonon mode, different to that observed in geometrically
frustrated ZnCr$_2$O$_4$ \cite{Sushkov}. As the AFM order in
ZnCr$_2$Se$_4$ can be shifted to $T = 0$~K in external fields of
approximately 6~T \cite{Hemberger06} also the mode splitting can be
suppressed in external magnetic fields resulting in spin-phonon
coupling typical for ferromagnets.

Single crystalline ZnCr$_2$Se$_4$ was grown by a liquid transport
method from ZnSe and CrCl$_3$ starting materials at temperatures
between 900 and 950$^\circ$C. X-ray diffraction on crushed single
crystals revealed a cubic single phase material with lattice
constant $a = 1.0498$~nm and selenium fractional coordinate $x =
0.385$. As shown earlier \cite{Hemberger06}, the magnetic
susceptibility can be described by a "ferromagnetic" CW temperature
of 90~K and an effective chromium moment which is very close to the
spin-only value of Cr$^{3+}$ in an octahedral environment (S=3/2).
The reflectivity experiments were carried out in the far infrared
range using the Bruker Fourier-transform spectrometer IFS113v
equipped with a He bath cryostat and a split-coil magnet for
external magnetic fields up to 7~T.
\begin{figure}[]
\includegraphics{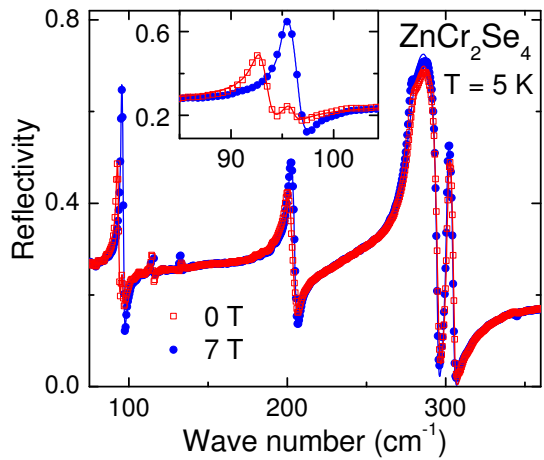}
\caption{(color online) Reflectivity vs. wave number of
ZnCr$_2$Se$_4$ at 5~K in zero magnetic field (open red squares) and
in an external magnetic field of 7~T (closed blue circles). The
small spikes close to 115~cm$^{-1}$ and 130~cm$^{-1}$ are
experimental artefacts. The inset shows the wave number regime close
to phonon mode 1, which reveals a splitting in zero field. The solid
lines correspond to fits as described in the text.} \label{fig1}
\end{figure}

Figure~\ref{fig1} shows the reflectivity spectra of ZnCr$_2$Se$_4$
in the range from 70 to 400 cm$^{-1}$ as measured at 5~K in external
magnetic fields of 0 and 7~T.  As expected from lattice symmetry,
the spectrum consists of four phonon triplets of F$_{1u}$ symmetry.
The room temperature spectra agree well with those previously
measured on ceramic samples \cite{Lutz, Wakamura76}. The spikes
between 130 and 170 cm$^{-1}$ are due to scattering processes and/or
resonances in the windows of the split-coil magnet and will not be
discussed further. The dielectric function $\epsilon(\omega)$ is
obtained by calculating the factorized form
\begin{equation}\label{dielconst}
\epsilon(\omega) = \epsilon_{\infty} \prod_j
\frac{\omega^2_{Lj}-\omega^2-i\gamma_{Lj}\omega}{\omega^2_{Tj}-\omega^2-i\gamma_{Tj}\omega}.
\end{equation}
Here $\omega_{Lj}$, $\omega_{Tj}$, $\gamma_{Lj}$ and $\gamma_{Tj}$
correspond to longitudinal (L) and transversal (T) eigenfrequency
($\omega_j$) and damping ($\gamma_j$) of mode $j$, respectively. At
normal incidence the real and imaginary part of the dielectric
function, $\epsilon$' and $\epsilon$'', are related to the
reflectivity via
\begin{equation}\label{R}
R = \frac{(n-1)^2+k^2}{(n+1)^2+k^2}
\end{equation}
and the relations $n^2 - k^2 = \epsilon$' and $2nk = \epsilon$''.
From the measured reflectivity the values of $\epsilon_{\infty}$,
$\omega_{Lj}$, $\omega_{Tj}$, $\gamma_{Lj}$ and $\gamma_{Tj}$ have
been determined using a fit routine developed by A. Kuzmenko
\cite{Kuzmenko}. Results of these fits are shown in Fig.~\ref{fig1},
namely at 0~T and 7~T, at a temperature of 5~K.

\begin{figure}[]
\includegraphics{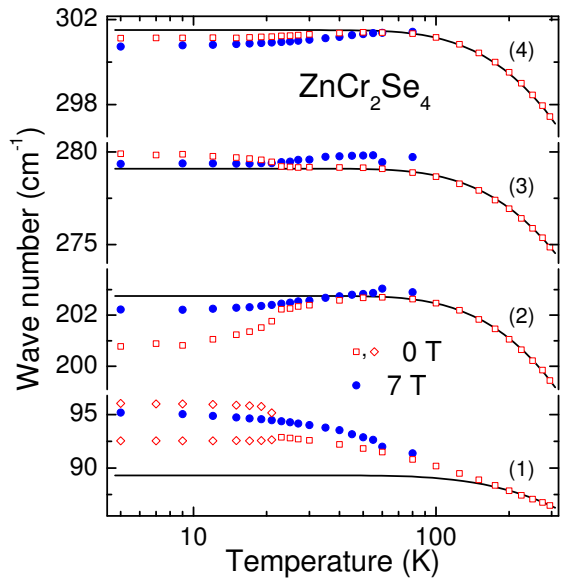}
\caption{(color online) Eigenfrequencies of the phonon modes in
ZnCr$_2$Se$_4$ as function of temperature. Eigenfrequencies at zero
external fields (open red squares/diamonds) are compared to those
observed at 7~T (closed blue circles). The solid lines are fits to
the high temperature results using a simple anharmonic term, Eq.
(3).} \label{fig2}
\end{figure}

The temperature dependence of the eigenfrequencies of the 4 IR
active modes of ZnCr$_2$Se$_4$ in zero external magnetic field is
shown in Fig.~\ref{fig2} (open red squares, and open red diamonds
for the split modes for $T < T_N$) on a semi-logarithmic plot. With
decreasing temperature the phonon eigenfrequencies increase as
usually observed in anharmonic crystals. To get an estimate of this
purely anharmonic contribution to the phonon modes, we fitted the
temperature dependence assuming
\begin{equation}\label{wfit}
\omega_j =
\omega_{0j}\left(1-\frac{c_j}{\mathrm{exp}(\Theta/T)-1}\right).
\end{equation}
Here $\omega_{0j}$ indicates the eigenfrequency of mode $j$ in the
absence of spin-phonon coupling at 0~K, $c_j$ is a mode dependent
scale factor of the anharmonic contributions and $\Theta = 309$~K a
rough estimate of the Debye temperature, as determined from an
arithmetic average of the IR active phonon frequencies. The results
of these fits to the high-temperature eigenfrequencies ($T > 100$~K)
are indicated as solid lines in Fig.~\ref{fig2}. It is clear that
all modes reveal significant deviations from this purely anharmonic
behavior for $T < 100$~K. The deviations are of the order of some
percent and are positive for mode 1 and mode 3, but negative for
modes 2 and 4. The anomalous temperature dependence smoothly evolves
below 100~K, a temperature scale corresponding to the Curie-Weiss
temperature $\Theta_{\mathrm{CW}}=90$~K, but significantly becomes
enhanced just below $T_\mathrm{N}$, indicating the temperature
dependence of an order parameter.

In addition to these anomalies in the temperature dependence, mode 1
reveals a clear splitting below the antiferromagnetic phase
transition at $T_\mathrm{N}=21$~K. These is exemplified in the inset
of Fig.~\ref{fig1}, were the results at 5~K in zero magnetic field
reveal a clear two-peak structure of the reflectivity. The splitting
amounts 3.5~cm$^{-1}$ and obviously results from the breaking of
cubic symmetry by the onset of antiferromagnetic order. Already at
this point it is worth mentioning that a similar splitting, however
of mode 2, has been observed in geometrically frustrated
ZnCr$_2$O$_4$ \cite{Hemberger97} and a splitting of all modes has
been observed in bond frustrated ZnCr$_2$S$_4$ \cite{Sushkov}. These
observations already signal the very different spin structures that
contribute to the symmetry-breaking perturbation of the dynamic
properties of these spinel compounds: The oxide experiences a
transition into a complex non-collinear spin structure, while the
sulfide undergoes a transition into a helical spin structure like
ZnCr$_2$Se$_4$, followed by a magnetic low-temperature state where
collinear and helical magnetic order coexist.

Eigenfrequencies, dielectric strength, and damping for all IR active
TO modes are listed in Tab.~\ref{tab1} for 5~K and 300~K. The
dielectric strength has been calculated via
\begin{equation}\label{Deps}
\epsilon(0)-\epsilon(\infty)=\sum_j\Delta\epsilon_j
=\epsilon_{\infty}\left(\prod_j\frac{\omega^2_{Lj}}{\omega^2_{Tj}}-1\right).
\end{equation}
In the case of non-overlapping modes, $\Delta\epsilon_j$ can
explicitly be derived:
\begin{equation}
\Delta\epsilon_j=\epsilon_{\infty}\left(\frac{\omega_{Lj}^2-\omega_{Tj}^2}{\omega_{Tj}^2}\right)\prod_{i=j+1}^n\frac{\omega_{Li}^2}{\omega_{Ti}^2},
\end{equation}
where $n$ is the total number of IR active phonon modes. At room
temperature the high-frequency dielectric constant has been
determined to be $\epsilon_{\infty}=8.1$ which compares reasonably
with published results \cite{Lutz, Wakamura76}.
\begin{table}\caption{\label{tab1}Eigenfrequencies $\omega_j$
[cm$^{-1}$], dielectric strength $\Delta\epsilon_j$ and damping
$\gamma_j$ [cm$^{-1}$] for the TO phonons in ZnCr$_2$Se$_4$ at 5~K
and 300~K in zero magnetic field.}
\begin{ruledtabular}
\begin{tabular}{ccccccc}
& \multicolumn{3}{c}{5~K}  & \multicolumn{3}{c}{300~K}\\
mode & $\omega_j$ & $\Delta\epsilon_j$ & $\gamma_j$ & $\omega_j$ & $\Delta\epsilon_j$ & $\gamma_j$\\
\hline
(1) & 92.6 & 0.23 & 1.1 & 86.5 & 0.32 & 3.8\\
    & 96.1 & 0.06 & 1.6 &       &      &     \\
(2) & 200.8 & 0.30 & 5.0 & 199.4 & 0.28 & 6.6\\
(3) & 279.9 & 0.94 & 8.0 & 274.9 & 0.93 & 6.8\\
(4) & 301.1 & 0.23 & 2.0 & 297.4 & 0.24 & 3.0\\
\end{tabular}\end{ruledtabular}
\end{table}
It is interesting to note that the dielectric strength does not
strongly depend on temperature when compared to related spinel
compounds and is unusually small for mode 4. The damping of modes~2
and 3 is strong, even at 5~K.
\begin{figure}[b]
\includegraphics{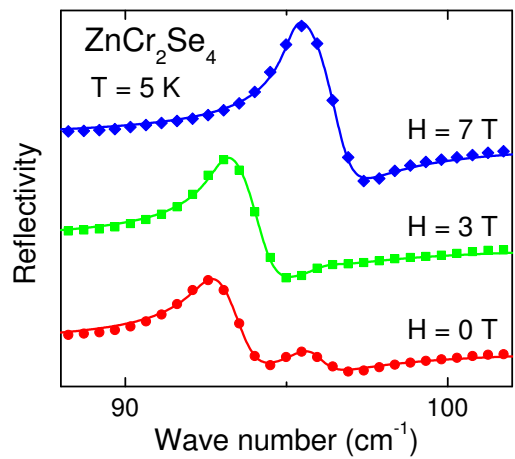}
\caption{(color online) Reflectivity vs. wave number for
ZnCr$_2$Se$_4$ around phonon mode 1 at 5~K and in magnetic fields of
0~T (red points), 3~T (green squares) and 7~T (blue diamonds). The
splitting of the phonon mode becomes fully suppressed at 7~T. The
lines are fits as described in the text.} \label{fig3}
\end{figure}

\begin{figure}[]
\includegraphics{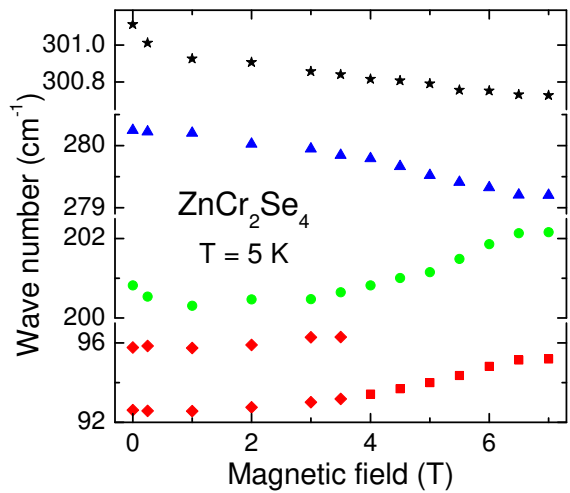}
\caption{(color online) Magnetic field dependence of the phonon
eigenfrequencies of ZnCr$_2$Se$_4$ at 5~K.} \label{fig4}
\end{figure}

The reflectivity at 5~K in an external magnetic field of 7~T is
shown in Fig.~\ref{fig1} and is compared with the reflectivity
obtained in zero external field. Minor changes only can be detected
for the high-frequency modes 3 and 4, while significant shifts are
obtained for phonons 1 and 2. But most spectacular, the splitting of
mode 1 becomes fully suppressed. This is documented in the inset of
Fig.~\ref{fig1}. We also followed the isothermal magnetic-field
dependence of all modes. As an example, Fig.~\ref{fig3} shows phonon
mode 1 at 5~K for three different fields of 0, 3 and 7~T. Already at
3~T the AFM-derived splitting can hardly be observed and vanishes
completely for 7~T. The dielectric function has been fitted for all
external fields and the calculated shift of the eigenfrequencies as
function of magnetic field is shown in Fig.~\ref{fig4}.
Figures~\ref{fig3} and \ref{fig4} provide experimental evidence that
the split branches of phonon 1 do not merge as function of
increasing magnetic field. The upper branch rather shows vanishing
intensity for magnetic fields $>$ 5 T. Modes 1 and 2 reveal an
increase, mode 3 and 4 a decrease of the eigenfrequencies on
increasing magnetic fields. This demonstrates that FM spin alignment
strengthens (weakens) the force constants responsible for the phonon
frequencies of mode 1 and 2 (3 and 4). The size of the overall
effect differs significantly for the different phonons and
approximately amounts 0.4, 0.3, - 0.15 and - 0.05~cm$^{-1}$/T for
modes 1, 2, 3, and 4 respectively.
Finally, in Fig.~\ref{fig2} we included the temperature dependence
of the eigenfrequencies for all IR active phonons as measured in
external magnetic fields of 7~T. Above approximately 90~K, which
corresponds to the CW temperature, magnetic field effects are
negligible. At low temperatures, a magnetic field of 6~T completely
suppresses the splitting of mode 1, increases the eigenfrequencies
for mode 2 but decreases the frequencies for modes 3 and 4. This
temperature dependence of the phonons at 7~T almost recovers the
situation in ferromagnetic CdCr$_2$S$_4$ \cite{Wakamura}. In this
compound the frequencies of modes 1 and 2 reveal positive shifts
when compared to normal anharmonic behavior, while modes 3 and 4
exhibit negative shifts. In ZnCr$_2$Se$_4$ only mode 2 deviates
significantly from this ferromagnetic-like behavior.  To explain the
results in the ferromagnetic Cr compound Wakamura followed the
arguments of Br\"{u}esch \textit{et al.}~\cite{Brueesch} and assumed
that modes 1 and 2 are mainly determined by force constants between
Cr and the anions, while modes 3 and 4 are mainly governed by A-site
cation-anion bonds. The former are believed to be determined by FM
90$^\circ$ exchange, the latter by AFM superexchange. These
arguments are compatible with our field dependent measurements. It
has been pointed out by Fennie and Rabe \cite{Fennie72} that the
patterns of eigendisplacements may be somewhat more complex. From ab
inito calculations they find that in CdCr$_2$S$_4$ for modes 1 and 2
the A-site cations move against both, the B-site cations and the
anions, explaining partly the low dielectric strength of these
modes.

In conclusion, we have investigated the spin-phonon coupling in
ZnCr$_2$Se$_4$ by IR spectroscopy. Below $T_\mathrm{N}$, we found a
splitting of the lowest frequency mode, different to the findings in
geometrically frustrated ZnCr$_2$O$_4$, where only the second mode
undergoes a splitting, and different to the observations in bond
frustrated ZnCr$_2$S$_4$, where all modes split. It seems evident
that the Cr spinels, which have a spherical charge distribution and
no spin-orbit coupling, are prone to exchange induced symmetry
breaking. But it is important to note that three different spinels,
geometrically frustrated ZnCr$_2$O$_4$, bond-frustrated
ZnCr$_2$S$_4$ and ZnCr$_2$Se$_4$, which is dominated by
ferromagnetic exchange, undergo spin-driven Jahn-Teller effects with
different patterns of phonon splittings. ZnCr$_2$Se$_4$ is close to
a ferromagnetic phase boundary and, to our knowledge, represents the
first example where a spin-driven Jahn-Teller effect can be
suppressed in external magnetic fields.

This work has been supported by the Deutsche Forschungsgemeinschaft
(DFG) via the collaborative research center SFB~484 (Augsburg).

\end{document}